\documentclass[conference]{IEEEtran}

\usepackage[utf8]{inputenc}
\usepackage{graphicx}
\usepackage{url}
\usepackage{hyperref}
\usepackage[frozencache,cachedir=.]{minted}

\usepackage{cite}
\usepackage{amsmath,amssymb,amsfonts}
\usepackage{algorithmic}
\usepackage{textcomp}
\usepackage{xcolor}
\def\BibTeX{{\rm B\kern-.05em{\sc i\kern-.025em b}\kern-.08em
    T\kern-.1667em\lower.7ex\hbox{E}\kern-.125emX}}

\newcommand{\cod}[1]{\texttt{#1}}
\newcommand{\mytitle}{Design and Reliability of a User Space Write-Ahead Log in Rust}

\usepackage{hyperref} 
\hypersetup{
     pdftitle={\mytitle}, 
     pdfauthor={Vitor K. F. Pellegatti, Gustavo M. D. Vieira},
     pdfdisplaydoctitle=true,
     hidelinks
}

\begin{document}

\title{\mytitle}

\author{\IEEEauthorblockN{Vitor K. F. Pellegatti}
\IEEEauthorblockA{Department of Computing at Sorocaba (DComp)\\
Federal University of São Carlos\\
Sorocaba, São Paulo, Brazil\\
Email: vitor.pellegatti@estudante.ufscar.br}
\and
\IEEEauthorblockN{Gustavo M. D. Vieira}
\IEEEauthorblockA{Department of Computing at Sorocaba (DComp)\\
Federal University of São Carlos\\
Sorocaba, São Paulo, Brazil\\
Email: gdvieira@ufscar.br}
}

\maketitle

\begin{abstract}
  Write-ahead logs (WALs) are  a fundamental fault-tolerance technique
  found  in many  areas of  computer science.   WALs must  be reliable
  while maintaining  high performance, because all  operations will be
  written to the WAL to  ensure their stability. Without reliability a
  WAL  is useless,  because  its utility  is tied  to  its ability  to
  recover  data  after a  failure.   In  this  paper we  describe  our
  experience creating a prototype user  space WAL in Rust. We observed
  that  Rust is  easy to  use,  compact and  has  a very  rich set  of
  libraries.  More  importantly, we  have found  that the  overhead is
  minimal, with the WAL prototype  operating at basically the expected
  performance of the stable memory device.
\end{abstract}

\begin{IEEEkeywords}
Fault tolerance, write-ahead log, Rust.
\end{IEEEkeywords}

\section{Introduction}

Write-ahead logs  (WALs) are  a fundamental  fault-tolerance technique
found in  many areas of computer  science~\cite{peterson1983log}, from
database         management        systems~\cite{jhingran1992analysis,
  haubenschild2020rethinking}  to  operating systems  and  distributed
systems~\cite{nakamura2019integration}. The  idea behind  a WAL  is to
commit  to stable  storage a  log  of operations  that \emph{will}  be
performed  in the  system,  before the  operations  are even  started.
Usually, the operations  logged are long and/or complex  and can leave
the  system in  an  invalid state  if interrupted  by  a fault  during
execution. By  consulting the log  during recovery after a  crash, the
system can  ensure these  operations are  effectively applied  and the
system stays consistent.

As  a central  piece  in  the fault  tolerance  puzzle,  WALs must  be
reliable while  maintaining high  performance, because  all operations
will be  written to  the WAL.  Without reliability  a WAL  is useless,
because its  utility is tied  to its ability  to recover data  after a
failure.   Moreover,  recovery  performance is  also  very  important,
because a  system under recovery  is an unavailable system.  To muddle
the waters further,  with the rising use of solid  state stable memory
such  as  SSDs,  reliability  of  the  storage  device  must  also  be
managed~\cite{zheng2016reliability}.

We see as  an important research subject the construction  of safe and
reliable  system-level infrastructure.   In this  context, the  widely
loved  Rust  language~\cite{klabnik2023rust}  offers  safer  code  and
C-level performance.   Safety in this  case usually refers  to correct
memory access, but  Rust provides more with a type  system that allows
safe concurrent  programming~\cite{jung2021safe} and  advanced program
analysis~\cite{balasubramanian2017system}.

In this paper  we describe our experience  creating a \emph{prototype}
user space WAL in Rust. Our  motivation was to assess if this language
has: 1) the abstractions to create a productive application level API,
2) the infrastructure to handle  serialization and low level file I/O,
3) actual reliability regarding crashes and file corruption and 4) low
overhead  leading to  good performance.   To assess  these points,  we
subjected our  prototype to  a series  of reliability  and performance
experiments.

We have  observed that  programming interfaces  and ergonomics  of the
language are  excellent, producing  very readable  code. Serialization
and file  I/O are  well supported  and mature, with  a good  number of
ready  made solutions  and  proper interfaces  for  creating our  own.
Reliability at the code level is excellent, but overall reliability is
limited by the on-disk  serialization format. Finally, our experiments
have  shown  that overhead  is  minimal,  operating at  basically  the
expected performance of the stable memory device.

This  paper  is  organized  as  follows.  Sections~\ref{sec:logs}  and
\ref{sec:rust}  give a  background in  WALs and  the Rust  programming
language. Section~\ref{sec:design}  describes the  design requirements
and  decisions  of  our WAL  prototype.   Section~\ref{sec:evaluation}
shows our experimental results and Section~\ref{sec:conclusion} closes
the paper with our conclusions.

\section{Write-Ahead Logs}
\label{sec:logs}

\emph{Write-ahead log}  (WAL) is  a recovery  technique first  used in
database managements  systems~\cite{peterson1983log}.  It  consists in
writing to  stable storage  a log  of changes to  be performed  in the
application state before effectively applying them. By forcing the log
write to be stable before performing  the state change, it is possible
to  recover  the  application  state afterwards  with  an  appropriate
recovery   algorithm~\cite{mohan1992aries}.    This  is   particularly
relevant  if the  state change  to be  performed in  composed of  many
writes  to  stable  memory  that  must be  atomic  to  maintain  state
consistency~\cite{jhingran1992analysis}.   A  \emph{checkpoint}  is  a
record in  the log  where the application  state itself  is completely
written to stable  storage and is used by many  recovery algorithms as
their starting point~\cite{haubenschild2020rethinking}.

Another simpler,  perhaps more interesting, way  that write-ahead logs
can be used is when \emph{no writes} are done to stable memory as part
of a state  change~\cite{vieira08a}. The application state  is kept in
main-memory and the  write-ahead log becomes the sole  registry of the
state change  in stable memory,  while the change happens  in volatile
memory only.   If a crash  occurs, the recovery algorithm  consists in
replaying all state  changes recorded in the log. We  call such WAL an
\emph{operation replay log}.  This  approach streamlines all writes to
stable memory  in sequential units  that extract the  most performance
out  of  the   stable  memory  subsystem~\cite{mingardi19}.  Recovery,
however, becomes very expensive, requiring reading all the log records
and reapplying them sequentially.  In this context, checkpoints in the
log are  very important to speed  up recovery and must  be paired with
dumps of the application state to stable memory~\cite{vieira08a}.

Write-ahead  logs are  a  very important  technique,  used in  several
contexts from databases to file systems. Research in WALs followed its
applications,  focused in  their  performance. For  example, the  most
critical step  for performance in a  WAL scheme is the  writing of the
individual records.   The write performance  can be improved  by using
parallel                      processing~\cite{johnson2012scalability,
  nakamura2019integration},  by  writing  to  novel  types  of  stable
memory~\cite{kim2016nvwal}  or  by  storing  the  WAL  in  a  distinct
device~\cite{rocha2012},   among  many   other  strategies.    Another
important     approach    to     improve    WAL     performance    are
\emph{log-structured} data structures, where  the application state is
kept      in       persistent      memory       \emph{inside}      the
WAL~\cite{sauer2018fineline}.  This  way, stable writes  are minimized
and recovery speed can be greatly increased.

Another research concern about WALs  is their reliability, as they are
central to recovery of applications in the presence of serious faults.
Some research  went into  securing the  WAL from  an attack,  to avoid
leaking  sensitive  information~\cite{pei2021bigfoot}.   However,  one
overlooked area is the integrity of  data once it is written to stable
storage.   By definition,  one never  expects data  written to  stable
storage to  change or otherwise  be corrupted, but  unfortunately this
can    happen,    specially    in     the    event    of    a    power
loss~\cite{zheng2016reliability}.  One  change in a record  in the WAL
can hinder recovery, or make it outright impossible.

\section{Rust}
\label{sec:rust}

The  Rust  programming  language~\cite{klabnik2023rust}  is  a  modern
alternative for  building systems  software, with  a primary  focus on
safe  programming. This  language,  through its  unique memory  model,
allows for  the compilation  of safe code  into binaries  that execute
with minimal overhead.  These programs combine performance on par with
C/C++ with  the memory safety of  application languages. Additionally,
Rust is a modern language, with tools and libraries that allow for the
same  level  of  productivity achieved  with  application  programming
languages.

Rust memory consistency mechanism makes  it an extremely safe language
in terms of memory access, without significant overhead at runtime and
without  a garbage  collector.  Some  research also  shows interesting
properties  in Rust  code, such  as software  fault isolation,  static
information  flow analysis  control, and  easily saving  the program's
execution state~\cite{balasubramanian2017system}.

The main  mechanism that allows  the language to  be safe and  fast is
called \emph{ownership}.   In Rust, variable declarations  are made at
compile time,  and each variable is  intrinsically tied to a  piece of
memory allocated to it.  These two assumptions ensure that the program
knows at  all times which variable  is responsible for which  piece of
memory, preventing undefined behavior  in the code.  Additionally, the
language allows references to be  created, and there is strict control
over  which variable  can alter  which piece  of memory  at all  times
during code execution.

Besides its memory safety mechanisms of ownership and references, Rust
also has many  of the features found in modern  languages. Among them,
we can mention \emph{traits}  and \emph{derive}, two essential aspects
for a  modern WAL  implementation.  Traits  are similar  to interfaces
found in other programming languages, allowing for easy replication of
behavior across  different structures,  albeit with  some differences.
Derive is a call to the compiler that enables metaprogramming and code
addition during compilation.  Through derive,  we can ask the compiler
to provide  us with  basic implementations of  certain traits  for any
classes in the language.

A WAL  implementation written in  Rust can  be extremely generic  as a
consequence of these two  important language features.  Anyone wishing
to store a record in the WAL, could use the derive macro on the struct
that defines  the record.  This  will allow automatic  serialization e
deserialization of the struct. Combining this with the definition of a
trait  that  defines   minimum  functionality  of  a   record,  it  is
straightforward to  create an API  to write and retrieve  records from
stable memory.

\section{Design of a Write-Ahead Log in Rust}
\label{sec:design}

We  created  an  operation  replay  log that  can  be  used  to  store
application  state changes,  and later  replay them  if necessary  for
recovery.  In this section we  describe our requirements for this WAL,
design decisions and its implementation.

\subsection{Requirements and Assumptions}

The  initial requirement  for the  WAL  created is  that it  is to  be
general purpose,  in the sense  that a record  stored isn't tied  to a
specific format from the point of view of the WAL. Put in another way,
the WAL  treats records as  opaque data  structures and does  not care
about their contents. Moreover, we want a system with a high degree of
generality,  capable of  accepting any  Rust data  structure and  then
translating it into the desired serialization formats.  As the records
are opaque, the  system relies in the  serialization infrastructure of
the Rust language.  This allows for  the actual log file to be written
in several formats, as discussed later.

The  system  offers several  key  operations.   It creates  WAL  files
whenever requested  by the application,  and makes them  available for
use.   It   efficiently  writes  records  to   the  appropriate  files
\emph{synchronously}, ensuring stability of  written records.  It also
efficiently  recovers written  records,  with  minimal overhead.   The
recovery process provides the application  with an ordered list of all
records  in  the log,  that  it  can  process  to recover  its  state.
Furthermore, it ensures the integrity of records if the writing format
allows for such verification.

We've designed the WAL to always  be available while in use, much like
an open  file. This  behavior is  maintained even  in the  presence of
failures. For writes, this means  alerting the client application that
the record wasn't written and therefore isn't stable.  The application
can then rewrite the record until  it succeeds. For reads, the systems
interrupts the recovery process as soon as an error is encountered and
the client is  alerted of issues when recovering  records.  This means
that once it's  started, the recovery process will  provide the client
application with all stored records, either  until the end of the file
or until a corrupted record is found.

This  recovery behavior  is  driven by  the fact  that  the WAL  would
primarily function as  a replay mechanism. The records in  the WAL are
written in order and reflect the succession of state changes. Thus, to
precisely  recreate the  system's  conditions before  a failure,  more
recent events  intrinsically rely on  previous ones. It  wouldn't make
sense to  recover objects that came  after a corrupted object,  as the
entire recovery would already be compromised.

\subsection{Application Programming Interface}

The first step in implementing the design shown in the last section is
the definition of  a suitable API (application  program interface). We
decided to design the simplest API that would satisfy the requirements
to avoid  adding confounding  factors to the  experimental evaluation.
In brief,  the WAL  API is  defined with  these three  simple abstract
operations:

\begin{LaTeXdescription}
\item[\cod{new(file) -> wal:}] Creates  a new WAL instance for
  logging data to the specified file.
\item[\cod{write(record):}] Writes a single record to the WAL.
\item[\cod{retrieve() ->  iterator:}]  Retrieves an  iterator
  over the records present in the WAL, in the order they were written.
\end{LaTeXdescription}

The type of  the records are defined as generic  types, constrained by
the  traits  \cod{Serialize}  for writing  and  \cod{Deserialize}  for
reading. These  traits are defined  in the popular  Rust serialization
framework Serde\footnote{\url{https://serde.rs/}},  that also provides
the   backend    implementation   of    the   reading    and   writing
process.  Concerning  the  client  application,  the  programmer  must
annotate     the      type     of     the     record      with     the
\verb|#[derive(Deserialize, Serialize)]| tag, easily obtaining default
implementations of both traits if  using regular Rust types. Moreover,
because it  uses statically defined  generic types, the WAL  will only
write and read a single type of record per file.

Using the Serde  framework we actually defined  two implementations of
the WAL, each based upon an  in disk format: JSONLogger and BinLogger.
The JSONLogger implementation  stores the records in  the JSON format,
while the  BinLogger implementation stores the  records in MessagePack
format\footnote{\url{https://msgpack.org/}}.    Each   implementation
provides  the API  with methods  of the  structs \cod{JSONLogger}  and
\cod{BinLogger}.  The  \cod{new()} method  is  static  and returns  an
implementation of WAL with the desired disk format. Both \cod{write()}
and \cod{retrieve()} are instance methods, and must be called from the
returned WAL instance returned.

\begin{listing}[!ht]
\begin{minted}[fontsize=\footnotesize,xleftmargin=18pt,linenos]{rust}
use logging_system::BinLogger;
use serde::{Deserialize, Serialize};

#[derive(Deserialize, Serialize)]
struct Record {
    id: u32,
}

fn main() {
    let logger: BinLogger<Record> =
        BinLogger::new("records.wal");
    
    let record = Record {
        id: 42,
    };

    if let Err(e) = logger.write_data(&record) {
        println!("Something went wrong: {e}");
    };

    let records =
        logger.retrieve_iterator().unwrap();
    for record in records {
        // Restore record
    }
}
\end{minted}
\caption{API Example}
\label{listing:api}
\end{listing}

Listing~\ref{listing:api}  shows   a  brief  example  with   the  main
components  of  the WAL  API  using  BinLogger.  After  the  necessary
imports, we  define the record data  structure in Lines 4  to 7, using
Serde's \cod{Deserialize}  and \cod{Serialize} traits for  derive. The
WAL is instantiated in Lines 10 and 11, and a sample record is written
in Line  17. Lines  21 to  25 show  how the  records can  be recovered
through an iterator.

\subsection{Internals}

The  WAL  system implementation  is  quite  straightforward, but  some
design decisions have implications  to the reliability and performance
of   the   system.    The   main   struct   of   each   implementation
(\cod{JSONLogger}   and   \cod{BinLogger})   are  created   with   the
\cod{new()} method.  In this function, we either create the file if it
doesn't exist  or link the structure  to an existing file  and keep it
open as long as the structure  is in use.  This approach boosts system
performance by  eliminating the need  to open  and close the  file for
every record written.

The two  main API primitives  are \cod{write()} and  \cod{read()}. The
implementation of \cod{write()} is mostly delegated to Serde, thus the
system design  consisted basically  on selecting  the disk  format and
adding  reliability. The  implementation of  \cod{read()} bridges  the
stream of data produced by Serde in the form of an iterator of records
that can be used for recovery.

\subsubsection{\cod{write()}}

The Serde framework provides many serialization formats, many of these
are well-known  text formats,  like JSON, YAML,  TOML, and  CSV, while
others are more specific to Rust,  such as RON (Rust Object Notation).
Serde also supports  binary formats like CBOR,  BSON, and MessagePack,
among many others.   For this work, we selected two  of these formats:
JSON and MessagePack.

JSON (JavaScript  Object Notation)  is particularly  appealing because
it's  a widely  recognized text  format used  in many  other contexts.
Moreover,  its  objects  are   extremely  generic,  which  was  highly
desirable  considering our  design  requirement  of generality.   This
provides a  degree of  interoperability with other  systems and,  as a
human readable format, makes development and testing easier.

MessagePack is relatively  newer than JSON.  Its core idea  is to be a
more  concise version  of JSON,  yet just  as efficient  or even  more
so. The  central characteristic of this  format is that it's  a binary
serialization  format, meaning  data  gets converted  into raw  bytes,
which  makes serialized  data very  compact. However,  reading objects
from a log  file is a challenge  and data corruption can  be a greater
problem than with JSON.

\begin{figure}[htbp]
\footnotesize
\begin{verbatim}
JSON 27 bytes
{"compact":true,"schema":0}
\end{verbatim}

\begin{verbatim}
MessagePack 18 bytes
82a7 636f 6d70 6163 74c3 a673 6368 656d 6100
\end{verbatim}
\caption{JSON and MessagePack}
\label{fig:mp-object}
\end{figure}

In Figure~\ref{fig:mp-object}, you  can see a small example  of a JSON
object   with    two   fields    and   its    equivalent   MessagePack
representation. The format uses bytes  to identify upcoming fields and
their respective  sizes. The MessagePack format  proved interesting to
our research because of its space efficiency and potential performance
gains related to having to write less data.  Also, this format is good
for assessing the  reliability of the WAL, because the  format is more
sensitive to  data corruption.  After  selecting the disk  formats, we
used the appropriate Serde data formats implemented in the Rust crates
(libraries) serde\_json\footnote{https://crates.io/crates/serde\_json}
and rmp-serde\footnote{https://crates.io/crates/rmp-serde}.

Both serialization formats possess distinct characteristics concerning
reliability and  size. To improve  the reliability of  the MessagePack
format we decided  to add a simple checksum to  the serialized record.
Neither Serde or the MessagePack format have support for checksumming,
so we added it with a simple strategy. First we use Serde to serialize
the record in  a byte array, then we checksum  the array and serialize
the checksum in a second byte array. We then write to the WAL file the
pair  of  serialized  record  and serialized  checksum.   As  checksum
algorithm we selected the simple and fast CRC32 algorithm, implemented
by                            the                            crc32fast
crate\footnote{\url{https://crates.io/crates/crc32fast/1.4.0}}. 

\subsubsection{\cod{read()}}

The implementation  of \cod{read()}  must perform  the deserialization
process, reverting the serialization.  From our requirements, we don't
need to read an arbitrary position in  the WAL, but only to stream the
records in  order since the  last checkpoint, to feed  the application
recovery process.   Thus, considering the Rust  language usual idioms,
the perfect fit for reading the records in an iterator. Rust iterators
provide   an   idiomatic   way   of  processing   a   list   of   data
\emph{efficiently},  by only  realizing each  list elements  lazily as
they are required.  This is desirable as it won't be necessary to read
the entire WAL file to begin processing it, saving time and memory.

The actual  implementation of the  reading takes the  reading iterator
provided by Serde  and transforms it in a  checksum validated iterator
of  records. This  is done  by  reading the  pairs (record,  checksum)
previously written  to the WAL  file, validating the  checksum against
the restored record,  and returning the record only if  it matches the
expected checksum.  We must note that  the WAL files created are bound
to  the record  type  written  within them.   Reading  something of  a
different type,  by providing  the wrong file  name for  example, will
cause deserialization errors. This is  done using the generic types of
Rust  and effectively  means WAL  files  have their  type enforced  at
compile time.

\section{Evaluation}
\label{sec:evaluation}

We created  a testing  environment to asses  both the  reliability and
performance of  our WAL  system.  First we  have a  simple application
that writes a set of records in the WAL and later recovers the written
records. Each record  is a small but complex Rust  struct, composed by
some primitives  fields and  a vector of  another struct  of primitive
fields.

\begin{listing}[!ht]
\begin{minted}[fontsize=\footnotesize,xleftmargin=18pt,linenos]{rust}
#[derive(Deserialize, Serialize, Debug)]
struct Data {
    a: u32,
    b: u32
}

#[derive(Deserialize, Serialize, Debug)]
struct Record {
    id: u32,
    comment: String,
    objects: Vec<Data>,
}
\end{minted}
\caption{Record Struct}
\label{listing:data}
\end{listing}

The  application  was  run  with  both  implementations  of  the  WAL:
JSONLogger and BinLogger.   The chosen record type  is directly mapped
to  the  two formats  without  further  intervention. The  reliability
experiment  was made  by manually  corrupting  the WAL  file on  disk,
before it  was recovered.  The  performance experiment consisted  in a
micro-benchmark  where  a  set  of  records  were  written  and  later
recovered from the WAL.

\subsection{Reliability}

We present  the reliability experiments by  serialization format.  For
the JSON  format, we corrupted  the WAL  file considering the  way the
JSON format is structured. We now list each failure introduced and the
consequence of each:

\begin{LaTeXdescription}
\item[Adding  or  removing spaces,  newlines,  or  tabs:] This  change
  doesn't  affect  the  object's  content  at  all,  only  the  file's
  formatting,  which  has  no  effect on  the  object  deserialization
  process.
  
\item[Adding or removing structural symbols:] Inserting any loose
  structural symbols of  a JSON object (\verb|{,},[,]|) will cause
  deserialization to fail at that specific point.
  
\item[Altering, removing,  or adding valid fields:]  Adding new fields
  didn't  alter  the object  recovery  process  in any  way.  However,
  removing  or  changing   field  identifiers  caused  deserialization
  failures for the altered object.
  
\item[Changing the  order of  fields:] This  change didn't  affect the
  object  recovery process.   The  deserialization  process is  robust
  enough to handle  both unnecessary additional fields  and changes in
  their order.

\item[Introducing  an empty  object:] Introducing  a completely  empty
  object  led  to  deserialization   problems  where  the  object  was
  inserted. As  an empty object  doesn't possess the  necessary fields
  for deserialization, this could be  considered a subcase of removing
  fields.

\item[Changing  values  of valid  fields:]  This  change represents  a
  corruption  that  the  system  cannot clearly  detect  in  the  JSON
  format. If  the altered value  is still consistent with  the field's
  type, the change won't be detected  as an error, and the record will
  be recovered normally, which is, in fact, an error.  However, if the
  value is  changed to a  different type  than expected, the  error is
  detected.
\end{LaTeXdescription}

In our evaluation the flexible nature of the text based format such as
JSON is both a blessing and a curse. The WAL file in this format isn't
vulnerable to some of the  corruptions performed, allowing the records
to be recovered.  Some corruptions in key areas of the file will still
render  it unreadable,  but that's  expected.  More  troubling is  the
observation that  some types  of corruption  will actually  change the
records stored, but these changes won't be noticed.

Considering  this limitation  of  the JSON  format,  and the  inherent
fragility of a  binary only format, we have added  checksumming to the
MessagePack serialization  format.  This  changes the  failure dynamic
completely,  because  now  any  and   all  changes  will  violate  the
checksum.  To assess this, we once again corrupted the WAL file, now
considering the way the MessagePack format is structured:
\begin{itemize}
\item Changes to an object's identifier;
\item Changes to a field's identifier;
\item Changes to a field's value;
\item Changes to the checksum identifier;
\item Changes to the checksum value.
\end{itemize}

For all  five items,  as expected,  our checksumming  procedure proved
robust enough to detect the  mentioned errors.  It's important to note
that this detection is performed  individually for each object.  If an
error  is found,  the deserialization  process is  interrupted at  the
correct point.

Finally,  it's  worth mentioning  that  it's  not always  possible  to
specifically identify  which change occurred.  For  instance, a change
to the  checksum value or an  object field's value triggered  the same
type  of  error.   However,  with  more  robust  error  detection  and
correction codes,  it might be  possible to detect the  specific error
introduced into a record.

\subsection{Performance}

To assess the WAL performance, we  ran an experiment focused on record
sizes.   As  we  described in  Section~\ref{sec:design},  each  record
written  to the  WAL must  be stable  in secondary  memory before  the
\cod{write()} call returns.  This  implies an expensive sync operation
after each write.   Thus, it is more efficient if  the application can
coalesce many records in a single  write.  This isn't always the case,
so we created some  scenarios reflecting changing application demands.
This also  characterizes the performance  expectation of the  WAL over
different record sizes.

The experiment consists  in writing a group of  2,000,000 records, and
later reading the records back. We created three configurations:
\begin{description}
\item[A:] 2,000,000 sets of 1 record;
\item[B:] 1,000,000 sets of 2 records;
\item[C:] 500,000 sets of 4 records.
\end{description}

The experiment was run in a host with  a AMD Ryzen 5 PRO 6650U CPU, 16
GB RAM,  and a  Kigston NV2 1TB  NVMe SSD. The  SSD has  a benchmarked
sequential  performance of  about 5,600  MB/s reading  and 2,800  MB/s
writing,  and a  random  access of  350 kIOPS  reading  and 300  kIOPS
writing. During  the experiment, CPU  and random access IOPS  were the
resources driving the results, SSD throughput was largely unused.

Each configuration was run 5 times for each implementation of the WAL:
JSONLogger and BinLogger. The  results were averaged by configuration,
separating reads  from writes,  using three  metrics: elapsed  time to
write all 2,000,000 records in seconds, throughput in MB/s, throughput
in records/s. Elapsed  time is an absolute metric  and summarizes well
the overall  performance of the  WAL, and the relative  performance of
each serialization  format. Throughput in bytes  shows how efficiently
each serialization format uses  the available disk throughput. Records
per  second shows  the  disk IOPS  utilization  of each  serialization
format.

\begin{table}[htbp]
  \caption{Write Performance Data}
  \label{tab:data:write}
  \centering
  \begin{tabular}{lccc} \hline
    & \multicolumn{3}{c}{Configurations} \\
    Metric & A & B & C\\ \hline
    JSON time (s) & 8.461 & 4.982 & 3.154 \\
    JSON throu. (MB/s) & 70.187 & 110.953 & 168.924 \\
    JSON throu. (rec./s) & 236389.85 & 401412.97 & 634115.41 \\ \hline
    MP time (s) & 7.179 & 3.858 & 2.171 \\
    MP throu. (MB/s) & 32.359 & 57.743 & 99.561 \\
    MP throu. (rec./s) & 278574.81 & 518376.44 & 921404.22\\ \hline
  \end{tabular}
\end{table}

Table~\ref{tab:data:write}  shows the  results  for write  operations.
The write  performance of the  WAL is  excellent, mostly bound  by the
maximum IOPS of the device, at about 250 k records per second for both
serialization formats.   Remember that every  write must be  synced to
stable  storage,  thus  the  latency of  individual  writes  dominates
here. The performance increases as we bunch more records together, and
this gives  an advantage to  the more compact MessagePack  format over
JSON, by  using less  CPU and  writing roughly 2/3  of the  data. This
efficiency makes the BinLogger score  an impressive 920 k records/s in
writes over  about 630  k records/s for  JSONLogger. Both  numbers are
rather good  in general, and it  is interesting to note  that even the
less efficient JSON format leaves most of the SSD bandwidth unused.

\begin{table}[htbp]
  \caption{Read Performance Data}
  \label{tab:data:read}
  \centering
  \begin{tabular}{lccc} \hline
    & \multicolumn{3}{c}{Configurations} \\
    Metric & A & B & C\\ \hline
    JSON time (s) & 5.938 & 5.714 & 5.423 \\
    JSON throu. (MB/s) & 99.997 & 96.740 & 98.242 \\
    JSON throu. (rec./s) & 336791.05 & 349993 & 368785.96 \\ \hline
    MP time (s) & 0.709 & 0.561 & 0.526 \\
    MP throu. (MB/s) & 327.858 & 396.977 & 410.695 \\
    MP throu. (rec./s) & 2822466.84 & 3563791.87 & 3800836.18 \\ \hline
  \end{tabular}
\end{table}

Table~\ref{tab:data:read} shows  the results for read  operations. The
read performance is mostly CPU bound, because there is no need to sync
data to and  from stable storage and, most importantly,  OS caches are
primed with the data just written  in the previous step. We considered
to bypass the  OS disk cache, but decided against  it because this way
we  could assess  the CPU  cost  of each  serialization format.   This
allowed  us  to  observe  that  MessagePack is  about  10  times  more
efficient than JSON  to deserialize data. Both formats  barely use any
disk bandwidth, and the impressive  number of 2,8 M records/s achieved
by the BinLogger,  while reading a record at a  time, clearly shows we
are not operating  bound by the SSD device  performance.  Both formats
improve their reading performance as the number of records in each set
increases, but only  by a small margin that we  hypothesize is related
to a smaller number of system calls.

\section{Conclusion}
\label{sec:conclusion}

We've built a  reliable WAL implementation in Rust  using two distinct
formats, each  offering different guarantees. The  first format, JSON,
is human-readable is robust against various file corruptions. However,
this  format  allows  the data  to  be  corrupted  in  a way  that  is
imperceptible  to   the  client   application.   The   second  format,
MessagePack, is a very fast binary format and combined with a checksum
mechanism    is    able    to     detect    and    flag    all    file
corruptions. Performance of both implementations were quite good, with
the most overhead  found in the JSON format.  MessagePack encoded logs
showed  a very  small overhead,  with  performance very  close to  the
maximum supported by the device.

The Rust programming language proved why  it is such a sensation among
system programmers. It is easy to use, compact and has a very rich set
of  libraries. We  were  able to  realize our  design  with almost  no
changes and used many  prepackaged components. Our reliability testing
found  a  very  safe  implementation,  able  to  detect  all  inserted
corruptions with the aid of a checksum.

Moving forward, we want to perform  a more thorough set of reliability
tests, and develop ways to recover a damaged WAL file. If the checksum
is  replaced by  an error  correcting code,  some level  of corruption
could  be   withstood.   We  could   also  pursue  an   approach  that
incorporates even more security into  our system, by use of encryption
for example.

\bibliographystyle{IEEEtran}
\bibliography{wal.rust}

\begin{thebibliography}{10}
\providecommand{\url}[1]{#1}
\csname url@samestyle\endcsname
\providecommand{\newblock}{\relax}
\providecommand{\bibinfo}[2]{#2}
\providecommand{\BIBentrySTDinterwordspacing}{\spaceskip=0pt\relax}
\providecommand{\BIBentryALTinterwordstretchfactor}{4}
\providecommand{\BIBentryALTinterwordspacing}{\spaceskip=\fontdimen2\font plus
\BIBentryALTinterwordstretchfactor\fontdimen3\font minus
  \fontdimen4\font\relax}
\providecommand{\BIBforeignlanguage}[2]{{%
\expandafter\ifx\csname l@#1\endcsname\relax
\typeout{** WARNING: IEEEtran.bst: No hyphenation pattern has been}%
\typeout{** loaded for the language `#1'. Using the pattern for}%
\typeout{** the default language instead.}%
\else
\language=\csname l@#1\endcsname
\fi
#2}}
\providecommand{\BIBdecl}{\relax}
\BIBdecl

\bibitem{peterson1983log}
R.~Peterson and J.~P. Strickland, ``Log write-ahead protocols and {IMS/VS}
  logging,'' in \emph{Proceedings of the 2nd ACM SIGACT-SIGMOD Symposium on
  Principles of Database Systems}, 1983, pp. 216--243.

\bibitem{jhingran1992analysis}
A.~Jhingran and P.~Khedkar, ``Analysis of recovery in a database system using a
  write-ahead log protocol,'' \emph{Acm Sigmod Record}, vol.~21, no.~2, pp.
  175--184, 1992.

\bibitem{haubenschild2020rethinking}
M.~Haubenschild, C.~Sauer, T.~Neumann, and V.~Leis, ``Rethinking logging,
  checkpoints, and recovery for high-performance storage engines,'' in
  \emph{Proceedings of the 2020 ACM SIGMOD International Conference on
  Management of Data}, 2020, pp. 877--892.

\bibitem{nakamura2019integration}
Y.~Nakamura, H.~Kawashima, and O.~Tatebe, ``Integration of {T}ic{T}oc
  concurrency control protocol with parallel write ahead logging protocol,''
  \emph{International Journal of Networking and Computing}, vol.~9, no.~2, pp.
  339--353, 2019.

\bibitem{zheng2016reliability}
M.~Zheng, J.~Tucek, F.~Qin, M.~Lillibridge, B.~W. Zhao, and E.~S. Yang,
  ``Reliability analysis of {SSD}s under power fault,'' \emph{ACM Transactions
  on Computer Systems (TOCS)}, vol.~34, no.~4, pp. 1--28, 2016.

\bibitem{klabnik2023rust}
S.~Klabnik and C.~Nichols, \emph{The {R}ust programming language}.\hskip 1em
  plus 0.5em minus 0.4em\relax No Starch Press, 2023.

\bibitem{jung2021safe}
R.~Jung, J.-H. Jourdan, R.~Krebbers, and D.~Dreyer, ``Safe systems programming
  in {R}ust,'' \emph{Communications of the ACM}, vol.~64, no.~4, pp. 144--152,
  2021.

\bibitem{balasubramanian2017system}
A.~Balasubramanian, M.~S. Baranowski, A.~Burtsev, A.~Panda, Z.~Rakamari{\'c},
  and L.~Ryzhyk, ``System programming in {R}ust: Beyond safety,'' in
  \emph{Proceedings of the 16th workshop on hot topics in operating systems},
  2017, pp. 156--161.

\bibitem{mohan1992aries}
C.~Mohan and F.~Levine, ``{ARIES/IM}: an efficient and high concurrency index
  management method using write-ahead logging,'' \emph{ACM Sigmod Record},
  vol.~21, no.~2, pp. 371--380, 1992.

\bibitem{vieira08a}
\BIBentryALTinterwordspacing
G.~M.~D. Vieira and L.~E. Buzato, ``{T}replica: Ubiquitous replication,'' in
  \emph{SBRC '08: Proc. of the 26th Brazilian Symposium on Computer Networks
  and Distributed Systems}, Rio de Janeiro, Brasil, May 2008. [Online].
  Available: \url{http://www.lbd.dcc.ufmg.br/bdbcomp/servlet/Trabalho?id=7450}
\BIBentrySTDinterwordspacing

\bibitem{mingardi19}
\BIBentryALTinterwordspacing
W.~Mingardi and G.~Vieira, ``Characterizing synchronous writes in stable memory
  devices,'' in \emph{Anais do XVIII Workshop em Desempenho de Sistemas
  Computacionais e de Comunicação}.\hskip 1em plus 0.5em minus 0.4em\relax
  Porto Alegre, RS, Brasil: SBC, 2019. [Online]. Available:
  \url{https://sol.sbc.org.br/index.php/wperformance/article/view/6458}
\BIBentrySTDinterwordspacing

\bibitem{johnson2012scalability}
R.~Johnson, I.~Pandis, R.~Stoica, M.~Athanassoulis, and A.~Ailamaki,
  ``Scalability of write-ahead logging on multicore and multisocket hardware,''
  \emph{The VLDB Journal}, vol.~21, pp. 239--263, 2012.

\bibitem{kim2016nvwal}
W.-H. Kim, J.~Kim, W.~Baek, B.~Nam, and Y.~Won, ``{NVWAL}: Exploiting {NVRAM}
  in write-ahead logging,'' \emph{ACM SIGPLAN Notices}, vol.~51, no.~4, pp.
  385--398, 2016.

\bibitem{rocha2012}
P.~E. Rocha and L.~C. Bona, ``Analyzing the performance of an externally
  journaled filesystem,'' in \emph{2012 Brazilian Symposium on Computing System
  Engineering}.\hskip 1em plus 0.5em minus 0.4em\relax IEEE, 2012, pp. 93--98.

\bibitem{sauer2018fineline}
C.~Sauer, G.~Graefe, and T.~H{\"a}rder, ``{FineLine}: log-structured
  transactional storage and recovery,'' \emph{Proceedings of the VLDB
  Endowment}, vol.~11, no.~13, pp. 2249--2262, 2018.

\bibitem{pei2021bigfoot}
J.~Pei and V.~Shmatikov, ``{BigFoot}: Exploiting and mitigating leakage in
  encrypted write-ahead logs,'' \emph{arXiv preprint arXiv:2111.09374}, 2021.

\end{thebibliography}

\end{document}